\documentstyle[twocolumn,graphicx,twoside,aps,prl,floats]{revtex}
\begin{document}
\graphicspath{{../Figures/}{.}}
\title{Multi-Phase Patterns in Periodically Forced Oscillatory Systems}
\author{Christian Elphick}
\address{Centro de Fisica No Lineal y Sistemas Complejos de Santiago,\\
 Casilla 17122, Santiago, Chile}
\author{Aric Hagberg\thanks{{\tt Email: aric@lanl.gov}}}
\address{Theoretical Division and Center for Nonlinear Studies, MSB284,\\
Los Alamos National Laboratory, Los Alamos, NM 87545}
\author{Ehud Meron\thanks{{\tt Email: ehud@bgumail.bgu.ac.il}}}
\address{The Jacob Blaustein Institute for Desert Research and the Physics
Department,\\
Ben-Gurion University, Sede Boker Campus 84990, Israel}
\date{\today}

\twocolumn[\hsize\textwidth\columnwidth\hsize\csname
@twocolumnfalse\endcsname
\maketitle

\begin{abstract}
Periodic forcing of an oscillatory system produces frequency locking
bands within which the system frequency is rationally related to the
forcing frequency. We study extended oscillatory systems that respond
to uniform periodic forcing at one quarter of the forcing frequency
(the 4:1 resonance). These systems possess four coexisting stable
states, corresponding to uniform oscillations with successive phase
shifts of $\pi/2$.  Using an amplitude equation approach near a Hopf
bifurcation to uniform oscillations, we study front solutions
connecting different phase states. These solutions divide into two
groups: $\pi$-fronts separating states with a phase shift of $\pi$ and
$\pi/2$-fronts separating states with a phase shift of $\pi/2$. We
find a new type of front instability where a stationary $\pi$-front
``decomposes'' into a pair of traveling $\pi/2$-fronts as the forcing
strength is decreased. The instability is degenerate for an amplitude
equation with cubic nonlinearities. At the instability point a
continuous family of pair solutions exists, consisting of
$\pi/2$-fronts separated by distances ranging from zero to
infinity. Quintic nonlinearities lift the degeneracy at the
instability point but do not change the basic nature of the
instability.  We conjecture the existence of similar instabilities in
higher $2n$:$1$ resonances ($n=3,4,..$) where stationary $\pi$-fronts
decompose into $n$ traveling $\pi/n$-fronts. The instabilities
designate transitions from stationary two-phase patterns to traveling
$2n$-phase patterns. As an example, we demonstrate with a numerical
solution the collapse of a four-phase spiral wave into a stationary
two-phase pattern as the forcing strength within the 4:1 resonance is
increased.

\end{abstract}
\vspace{-0.4cm}
\vskip2pc]

\section{Introduction}

\begin{figure}
\centering\includegraphics[width=3.0in]{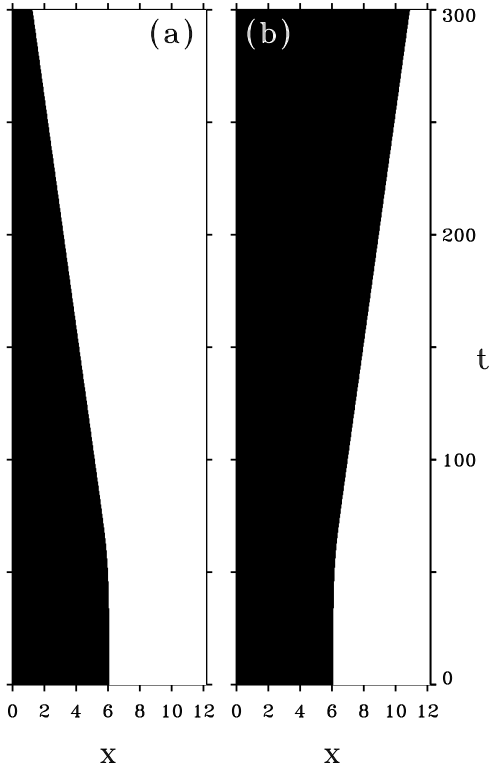}
\caption{
The NIB bifurcation in the 2:1 resonance: space-time plots showing 
an unstable stationary $\pi$-front (Ising)
evolving into left (a) 
and right (b) traveling $\pi$-fronts (Bloch) beyond the NIB
bifurcation. 
}
\label{nib}
\end{figure}

Periodic forcing of an oscillatory system produces a multiplicity of
uniform stable phase states. The simplest situation arises within the
2:1 frequency locking band where the system oscillates at one half of
the forcing frequency. In that case ``two-phase'' patterns appear,
involving alternating domains that oscillate with a phase shift of
$\pi$~\cite{Walg:97,CoEm:92,MaNe:94}. The boundaries
between nearby domains, hereafter $\pi$-fronts, may undergo a parity
breaking bifurcation rendering a stationary front unstable and giving
rise to a pair of counterpropagating fronts~\cite{CLHL:90}. This
instability, the so called ``nonequilibrium Ising-Bloch bifurcation''
(or NIB), designates a transition from standing two-phase patterns to
traveling two-phase patterns~\cite{HaMe:94a,BRSP:94,EHMM:97}.  The
instability is demonstrated in Fig.~\ref{nib} as a grey-scale map in
the space-time plane. Recent experiments on a photo-sensitive
Belousov-Zhabotinsky (BZ) reaction, periodically illuminated, have
also revealed a transition to labyrinthine patterns within the 2:1
band, suggesting the possible existence of a transverse instability of
$\pi $-fronts~\cite{POS:97}.

The situation becomes more complicated within the 4:1 band which has
four stable phase states shifted by $\pi/2$ with respect to one
another~\cite{Krau:95}. In addition to $\pi$-fronts there also exist
$\pi/2$-fronts separating oscillating domains with a phase shift of
$\pi/2$. The multiplicity of front solutions increases with the order
of the band. The 6:1 band has three types of fronts: $\pi$-fronts,
$2\pi/3$-fronts and $\pi/3$-fronts. The 8:1 band has four types of
fronts ($\pi$, $3\pi/4$, $\pi/2$ and $\pi/4$) and so on.  In addition
to adding new types of fronts as the band order is increased the
number of front solutions of a given type also increases.

\begin{figure}[h]
\centering\includegraphics[width=3.0in]{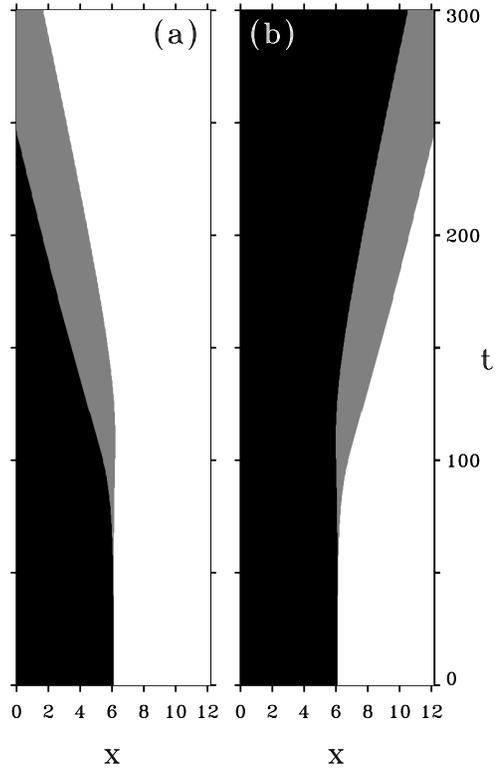}
\caption{The decomposition instability in the 4:1 resonance:
Space-time plots (solutions of Eqn.~(\protect\ref{fcglA}) showing the 
decomposition of an unstable $\pi$-front into a pair of $\pi/2$-fronts 
traveling to the left (a) or to the right (b). The
pairs of $\pi/2$-fronts enclose grey colored domains whose oscillation
phases are shifted by $\pi/2$ with respect to the black and white
domains.
Parameters in Eqn.~(\protect\ref{fcglA}): $\mu=1.0$, 
$\nu=0.02$, $\gamma_4=0.3$.  }
\label{decomposition}
\end{figure}

In this paper we report on a new instability of $\pi$-fronts,
occurring within the 4:1 band. Upon decreasing the forcing strength a
stationary $\pi$-front loses stability and decomposes into a pair of
traveling $\pi/2$ fronts. The instability is demonstrated in
Fig.~\ref{decomposition}.  The decomposition into a pair of traveling
$\pi/2$-fronts is accompanied by the appearance of an intermediate
(grey) domain whose phase of oscillation is shifted by $\pi/2$ with
respect to the adjacent white and black domains.  Like the NIB
bifurcation the $\pi$-front instability within the 4:1 band designates
a transition from stationary patterns to traveling waves. The
significant difference is that the two-phase stationary patterns give
place to traveling {\it four-phase} patterns. This feature of the 4:1
resonance is related to a peculiar property of the $\pi$-front
instability to be discussed in Section~\ref{instability}. The
$\pi$-front decomposition instability appears to exist in higher
$2n$:1 bands as well. We analyze in detail the 4:1 resonance case and
bring numerical evidence for the existence of this type of instability
in the 6:1 and 8:1 resonances. A brief account of some of the results
to be reported here has appeared in Ref.~\cite{EHM:98}.

We consider an extended system that is close to a Hopf bifurcation and
externally forced with a frequency about four times larger than the
Hopf frequency. The set of dynamical fields ${\bf u}$ describing the
spatio-temporal state of the system (e.g. set of concentrations in the
BZ reaction) can be written as ${\bf u}={\bf u_{0}}A\exp
{(i\frac{\omega _{f}}{4}t)} + c.c. + \dots$, where ${\bf u_{0}}$ is
constant, $A$ is a slowly varying complex amplitude, $\omega _{f}$ is
the forcing frequency and the ellipses denote smaller
contributions. The equation for the amplitude $A$ can be written in
the following standard form (after rescaling and shifting $\arg{A}$ by
a constant phase)~\cite{Newell:74,Gamb:85,EIT:87,CrHo:93}:
\begin{eqnarray}
A_{\tau} &=&(\mu +i\nu )A+(1+i\alpha )A_{zz}-(1-i\beta )|A|^{2}A
\nonumber \\ &&\mbox{}+\gamma _{4}{A^{\ast}}^{3}\,, \label{fcglA}
\end{eqnarray}
where the subscripts $\tau$ and $z$ denote partial derivatives with
respect to time and space, and all the parameters are real. The
proximity to the Hopf bifurcation implies $\mu <<1$. We will also be
using the following form of Eqn.~(\ref{fcglA}) obtained by rescaling
time space and amplitude as $t=\mu \tau$, $x=\sqrt{\mu/2}z$ and
$B=A/\sqrt{\mu}$:
\begin{eqnarray}
B_t &=&(1 +i\nu_0 )B+\frac{1}{2}(1+i\alpha )B_{xx}-(1-i\beta )|B|^{2}B
\nonumber \\ &&\mbox{}+\gamma _{4}{B^{\ast }}^{3}\,, \label{fcglB}
\end{eqnarray}
where $\nu_0=\nu/\mu$.

\section{Front solutions}

\begin{figure}
\centering\includegraphics[width=2.5in]{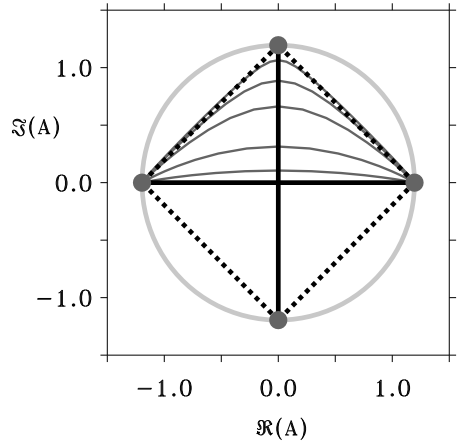}
\caption{
Phase portrait in the complex plane of solutions to
Eqn.~(\protect\ref{gl}).  The dots represent the 4 spatially uniform
phase locked solutions.  The solid lines are the $\pi$ front solutions
and the dashed lines are the $\pi/2$ fronts.  The thin lines in the
circle are the phase portrait of showing the collapse of a $\pi$ front
into two $\pi/2$ fronts.}
\label{4:1-phase}
\end{figure}

We first study the gradient version of Eqn.~(\ref{fcglB}) which is
obtained by setting $\nu_0=\alpha=\beta=0$:
\begin{equation}
B_t= B+\frac{1}{2}B_{xx}-|B|^2B +\gamma_4 {B^{*}}^3\,.  \label{gl}
\end{equation}
Eqn.~(\ref{gl}) has four stable phase states for $0<\gamma_4<1$ shown
by solid circles in Fig.~\ref{4:1-phase}: $B_{\pm 1}=\pm\lambda$ and
$B_{\pm i}=\pm i\lambda$, where $\lambda=1/\sqrt{{1-\gamma_4}}$. Front
solutions connecting pairs of these states divide into two groups,
$\pi$-fronts and $\pi/2$-fronts. The $\pi$-fronts, shown in
Fig.~\ref{4:1-phase} as solid lines, are given by
\begin{eqnarray}  \label{pifronts}
B_{-1\to +1}&=&B_{+1}\tanh{x}\,, \nonumber \\ B_{-i\to
+i}&=&B_{+i}\tanh{x}\,.
\end{eqnarray}
The $\pi/2$-fronts are shown in Fig.~\ref{4:1-phase} by the dashed
curves. For the particular parameter value $\gamma_4=1/3$ they have
the simple forms
\begin{eqnarray}  \label{pi/2fronts}
B_{+1\to
+i}&=&\frac{1}{2}\sqrt{\frac{3}{2}}\bigl[1+i-(1-i)\tanh{x}\bigr]\,,
\nonumber \\ B_{-i\to
+1}&=&\frac{1}{2}\sqrt{\frac{3}{2}}\bigl[1-i+(1+i)\tanh{x}\bigr]\,,
\nonumber \\ B_{+i\to -1}&=&-B_{-i\to +1}\,, \nonumber \\ B_{-1\to
-i}&=&-B_{+1\to +i}\,.
\end{eqnarray}
Additional front solutions follow from the invariance of
Eqn.~(\ref{gl}) under reflection, $x\to -x$. For example, the
symmetric counterparts of $B_{+i\to +1}(x)$ and $B_{+1\to -i}(x)$ are
$B_{+1\to +i}(x)=B_{+i\to +1}(-x)$ and $B_{-i\to +1}(x)=B_{+1\to
-i}(-x)$.

Consider now the nongradient system~(\ref{fcglB}). The main effect of
the nongradient terms is to make the $\pi/2$-fronts traveling. The
nongradient terms have no effect on the $\pi$-fronts which remain
stationary. To see this we assume a traveling solution $B(x-ct)$ of
Eqn.~(\ref{fcglB}) and project this equation on the translational mode
$B^\prime$. For $\pi$-fronts we obtain
\begin{equation}
c<{B_0^\prime}^2>=0 \qquad B_0(z)=\lambda\tanh{z}\,,
\end{equation}
implying $c=0$ (the brackets denote integration over the whole
line). For $\pi/2$-fronts with $\gamma_4=1/3$ we find
\begin{eqnarray}
\vert c\vert&=&\frac{\lambda}{<{B_0^\prime}^2>}\Bigl[(\nu_0+
\frac{1}{2}
\lambda^2\beta)<{B_0^\prime}>+\frac{1}{2}\beta<{B_0^2B_0^\prime}>
\Bigr] \nonumber \\ &=& \frac{3}{2}(\nu_0+\beta)\,, \label{pi2speed}
\end{eqnarray}
where $\lambda=\sqrt{3/2}$. A perturbation analysis around
$\gamma_4=1/3$ shows that the expression~(\ref{pi2speed}) for the
speed remains valid for small deviations of $\gamma_4$ from $1/3$.

\section{A $\protect\Pi$-front Instability}
\label{instability}
The $\pi$-fronts~(\ref{pifronts}) are similar to the Ising front in
the 2:1 band but as we will see shortly the instability they undergo
is not a pitchfork bifurcation like the NIB. It is rather a degenerate
instability leading to asymptotic solutions that are not smooth
continuations of the unstable stationary $\pi$-fronts in a sense to be
made clear in the following. A stability analysis of the $\pi$-fronts
indicates that they lose stability at $\gamma_4=1/3$. To analyze the
instability we study Eqn.~(\ref{fcglB}) near that critical value.

\subsection{Gradient system}

We begin with the gradient version~(\ref{gl}). Introducing the new variables 
\begin{equation}
U=\Re (B)+\Im (B)\qquad V=\Re (B)-\Im (B)\,, 
\end{equation}
we rewrite Eqn.~(\ref{gl}) as 
\begin{mathletters}
\label{UVeqn}
\begin{eqnarray}
U_{t} &=&U+\frac{1}{2}U_{xx}-\frac{2}{3}U^{3}-\frac{d}{2}(U^{2}-3V^{2})U\,, 
\label{Ueqn}   \\
V_{t} &=&V+\frac{1}{2}V_{xx}-\frac{2}{3}V^{3}-\frac{d}{2}(V^{2}-3U^{2})V\,,
\label{Veqn}
\end{eqnarray}
\end{mathletters}
where 
\[
d=\gamma _{4}-1/3\,.\nonumber
\]
At the instability point, $\gamma _{4}=1/3$, the two equations decouple
(since $d=0$) and admit solutions of the form 
\begin{eqnarray} 
\label{UVsol}
U &=&\sigma _{1}B_{0}(x-x_{1})\,, \nonumber \\
V &=&\sigma _{2}B_{0}(x-x_{2})\,,  
\end{eqnarray}
where $B_{0}(x)=\sqrt{\frac{3}{2}}\tanh {x}$, $\sigma _{1,2}=\pm 1$,
and $x_{1}$ and $x_{2}$ are arbitrary constants. An intuitive understanding of
this family of solutions can be obtained by expressing these solutions back
in terms of the complex amplitude $B$. For $\sigma _{1}=-\sigma _{2}=1$ for
example, the solution~(\ref{UVsol}) is equivalent to 
\[
B(x;x_{1},x_{2})=B_{-i\rightarrow +1}(x-x_{1})+B_{+1\rightarrow
+i}(x-x_{2})-\lambda \,. 
\]
When $|x_{2}-x_{1}|\rightarrow \infty $ this form approaches a pair of
isolated $\pi /2$-fronts: 
\[
B\approx B_{-i\rightarrow +1}(x-x_{1})\,,\qquad x\approx x_{1}\,, 
\]
and 
\[
B\approx B_{+1\rightarrow +i}(x-x_{2})\,,\qquad x\approx x_{2}\,. 
\]
When $x_{2}-x_{1}=0$ it reduces to the $\pi $-front $B_{-i\rightarrow +i}$.
Defining a ``center of mass'' coordinate, $\zeta $, and an order parameter,
$\chi $, by 
\[
\zeta =\frac{1}{2}(x_{1}+x_{2})\,,\qquad \chi =\frac{1}{2}(x_{2}-x_{1})\,, 
\]
the one-parameter family of
solutions, $\{\tilde B(x;\zeta ,\chi )~|~\chi \in R\}$, where 
$\tilde B(x;\zeta ,\chi )=B(x;x_{1},x_{2})$, represents 
$\pi/2 $-front pairs with distances, $2\chi $, ranging from zero to infinity.

For $\vert\gamma_4-1/3\vert=\vert d\vert\ll 1$, the weak coupling between the 
two equations~(\ref{Ueqn}) and~(\ref{Veqn}) induces slow drift along the solution family 
$B(x;x_{1},x_{2})$. We now write a pair solution as 
\begin{eqnarray}
U &=&\sigma _{1}B_{0}[x-x_{1}(t)]+u\,,  \nonumber \\
V &=&\sigma _{2}B_{0}[x-x_{2}(t)]+v\,,  \label{UVsolt}
\end{eqnarray}
where $u$ and $v$ are corrections of order $d$. Inserting these forms in
Eqns.~(\ref{UVeqn}) we obtain 
\begin{eqnarray}
{\cal H}_{1}u &=&\sigma _{1}\dot{x}_{1}B_{0}^{\prime }(x-x_{1})  
\label{H1} \\
&-&\frac{1}{2}d\sigma _{1}\bigl[B_{0}^{2}(x-x_{1})-3B_{0}^{2}(x-x_{2})\bigr]
B_{0}(x-x_{1})\,, \nonumber \\
{\cal H}_{2}v &=&\sigma _{2}\dot{x}_{2}B_{0}^{\prime }(x-x_{2}) \label{H2}\\
&-&\frac{1}{2}d\sigma _{2}\bigl[B_{0}^{2}(x-x_{2})-3B_{0}^{2}(x-x_{1})\bigr]
B_{0}(x-x_{2})\,,  \nonumber
\end{eqnarray}
where ${\cal H}_{1,2}=-1-\frac{1}{2}\frac{\partial ^{2}}{\partial x^{2}}
+2B_{0}^{2}(x-x_{1,2})$. Projecting the right hand side of Eqn.~(\ref{H1})
onto $B_{0}^{\prime }(x-x_{1})$, the zero eigenmode of 
${\cal H}_{1}^{\dagger }={\cal H}_{1}$, and setting to zero we obtain 
\begin{eqnarray}
\dot{x}_{1} &=&-\frac{27}{16}d\int_{-\infty }^{\infty }dx  \nonumber \\
&&\mbox{}\times \tanh (x-x_{1})~{\rm sech}^{2}(x-x_{1})~\tanh
^{2}(x-x_{2})\,,  \label{x1}
\end{eqnarray}
A similar solvability condition for~(\ref{H2}) leads to 
\begin{eqnarray}
\dot{x}_{2} &=&-\frac{27}{16}d\int_{-\infty }^{\infty }dx  \nonumber \\
&&\mbox{}\times \tanh (x-x_{2})~{\rm sech}^{2}(x-x_{2})~\tanh
^{2}(x-x_{1})\,.  \label{x2}
\end{eqnarray}
Expressing these equations in terms of $\zeta $ and $\chi $ we find 
\begin{eqnarray}
\dot{\zeta}=0\,,\label{zetadot} \\
\dot{\chi}=-\frac{27}{16}dJ(\chi )\,,
\label{chidot}
\end{eqnarray}
where 
\begin{equation}
J(\chi )=\int_{-\infty }^{\infty }dz~\tanh {z}~{\rm sech}^{2}{z}~\tanh
^{2}(z+2\chi )\,.\label{J}
\end{equation}
Evaluation of the integral in~(\ref{J}) yields 
\begin{eqnarray}
J(\chi ) &=&I(a)=6(a^{-1}-a^{-3})+(1-3a^{-2})G(a)\,,  \nonumber \\
G(a) &=&(1-a^{-2})\ln \Bigr(\frac{1+a}{1-a}\Bigl)\,,  \nonumber
\end{eqnarray}
where $a=\tanh {2\chi }$. Note that Eqns.~(\ref{zetadot}) and~(\ref{chidot}) 
are valid to {\em all}
orders in $\chi $ and to linear order around $\gamma _{4}=1/3$.

\begin{figure}
\centering\includegraphics[width=2.7in]{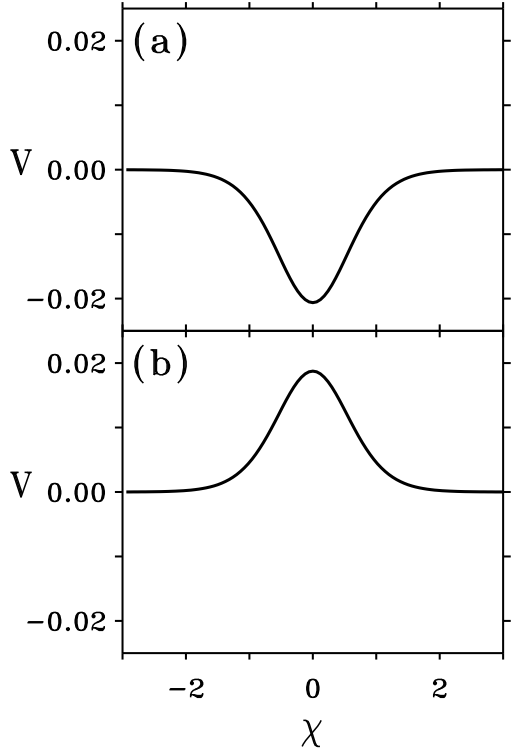}
\caption{The potential $V(\chi)$. (a) For $d>0$
the extremum at $\chi=0$ is a minimum and and $\chi$ converges to $0$.
(b) For $d<0$ the extremum is a maximum and $\chi$ diverges to $\pm
\infty$.  }
\label{pot}
\end{figure}

The equation for the order parameter~(\ref{chidot}) can be written in
the gradient form
\begin{equation}
\dot{\chi}=-\frac{dV}{d\chi}\,,\qquad V=\frac{27}{16}d\int^\chi J(z)
dz\,.
\end{equation}
Fig.~\ref{pot} shows the potential $V(\chi)$ for $d>0$
($\gamma_4>1/3$) and $d<0$.  There is only one extremum point,
$\chi=0$, of $V$. For $d>0$ it is a minimum and $\chi$ converges to
zero. Pairs of $\pi/2$-fronts with arbitrary initial separation,
$x_2-x_1$, attract one another and eventually collapse to a single
$\pi$-front ($x_1=x_2$ or $\chi=0$).  In practice, the collapse
process is noticeable only for relatively small separations.  For
$d<0$ the extremum point, $\chi=0$, is a maximum and $\chi$ diverges
to $\pm\infty$. A $\pi$-front decomposes into a pair of $\pi/2$-fronts
which repel one another. This process is shown in
Fig.~\ref{decomposition} for the nongradient system~(\ref{fcglA}). In
the gradient case both $\pi$ and$\pi/2$-fronts are stationary (in the
absence of interactions). Since the potential $V(\chi)$ becomes
practically flat at finite $\chi$ values, the pair of $\pi/2$-fronts
do not seem to depart from one another at long times.
Fig.~\ref{4:1-phase} shows the decomposition process of a $\pi$-front
in the complex $B$ plane. Starting with the $B_{-1\to+1}$ $\pi$-front,
represented by the thick solid phase portrait, the time evolution
(thin solid phase portraits) is toward the fixed point $B_{+i}$ and
the dashed phase portraits representing the pair of $\pi/2$-fronts
$B_{+1\to+i}$ and $B_{+i\to -1}$. Because of the parity symmetry
$\chi\to -\chi$, an appropriate perturbation of the initial
$B_{-1\to+1}$ $\pi$-front could have led the dynamics toward the pair
$B_{+1\to -i}$ and $B_{-i\to -1}$.  Notice that for $d=0$,
$\dot\zeta=0$, $\dot\chi=0$, and we recover the two-parameter family
of pair solutions $B(x;\zeta,\chi)$ with arbitrary $\zeta$ and
$\chi$. This degeneracy of solutions at $d=0$ is lifted by higher
order terms in the amplitude equation~(\ref{fcglA}) as will be
discussed in Section~\ref{ho} below.

\subsection{Nongradient system}

The results described above can easily be extended to the nongradient
system~(\ref{fcglB}) for small $\alpha$, $\beta$ and $\nu_0$. The
equations for $U$ and $V$ are
\begin{eqnarray}
U_{t}
&=&U+\frac{1}{2}U_{xx}-\frac{2}{3}U^{3}-\frac{d}{2}(U^{2}-3V^{2})U\,,
\nonumber \\ &&\mbox{}+\nu_0
V+\frac{\alpha}{2}V_{xx}+\frac{\beta}{2}(U^2+V^2)V\,, \nonumber \\
V_{t}
&=&V+\frac{1}{2}V_{xx}-\frac{2}{3}V^{3}-\frac{d}{2}(V^{2}-3U^{2})V\,,
\nonumber \\ &&\mbox{}-\nu_0
U-\frac{\alpha}{2}U_{xx}-\frac{\beta}{2}(U^2+V^2)U\,.
\label{UVeqs_ng}
\end{eqnarray}
Assuming $d$, $\alpha$, $\beta$ and $\nu_0$ are of the same order of
magnitude we write a solution in the form~(\ref{UVsolt}), insert in
Eqs.~(\ref{UVeqs_ng}) and obtain
\begin{eqnarray}
{\cal H}_{1}u &=&\sigma _{1}\dot{x}_{1}B_{0}^{\prime }(x-x_{1})
\nonumber \\ &-&\frac{1}{2}d\sigma
_{1}\bigl[B_{0}^{2}(x-x_{1})-3B_{0}^{2}(x-x_{2})\bigr]
B_{0}(x-x_{1})\nonumber \\ &+&\nu_0\sigma_2
B_0(x-x_2)+\frac{1}{2}\alpha\sigma_2
B_0^{\prime\prime}(x-x_2)\nonumber \\ &+&\frac{1}{2}\beta\sigma_2
\bigl[B_{0}^{2}(x-x_{1})+B_{0}^{2}(x-x_{2})\bigr] B_{0}(x-x_{2})\,,
\label{H1_ng}\\ {\cal H}_{2}v &=&\sigma _{2}\dot{x}_{2}B_{0}^{\prime
}(x-x_{2}) \nonumber \\ &-&\frac{1}{2}d\sigma
_{2}\bigl[B_{0}^{2}(x-x_{2})-3B_{0}^{2}(x-x_{1})\bigr]
B_{0}(x-x_{2})\nonumber \\ &-&\nu_0\sigma_1
B_0(x-x_1)-\frac{1}{2}\alpha\sigma_1
B_0^{\prime\prime}(x-x_1)\nonumber \\ &-&\frac{1}{2}\beta\sigma_1
\bigl[B_{0}^{2}(x-x_{1})+B_{0}^{2}(x-x_{2})\bigr] B_{0}(x-x_{1})\,.
\label{H2_ng}
\end{eqnarray}
Solvability conditions lead to equations for $\chi$ and $\zeta$. The
equation for $\chi$ remains unchanged. That is, Eqn.~(\ref{chidot}) is
valid for the nongradient equation~(\ref{fcglB}) as well. The equation
for $\zeta$ becomes
\begin{equation}
\sigma_1\sigma_2\dot\zeta= \nu_0 F_\nu(\chi)+\alpha
F_\alpha(\chi)+\beta F_\beta(\chi)\,,
\label{zetadot_ng}
\end{equation} 
where
\begin{eqnarray}
F_\nu&=&-\frac{3}{4}G(a)-\frac{3}{2}a^{-1}\,,\nonumber\\
F_\alpha&=&\frac{3}{4}I(a)\,,\nonumber\\
F_\beta&=&3a^{-1}\bigl(1-\frac{3}{2}a^{-2}\bigr)-\frac{9}{4}a^{-2}G(a)\,.
\nonumber
\end{eqnarray}
Notice that $F_\nu$, $F_\alpha$, and $F_\beta$ are odd functions of
$\chi$ and do not vanish when $d=0$. When $\vert\chi\vert\to\infty$
the right hand side of~(\ref{zetadot_ng}) converges to
$\frac{3}{2}(\nu_0+\beta)$, the speed of a $\pi/2$-front solution of
Eqn.~(\ref{fcglB}). The odd symmetries of $F_\nu$, $F_\alpha$, and
$F_\beta$ imply that the $\chi=0$ solution (representing a
$\pi$-front) remains stationary ($\dot\zeta=0$) in the nongradient
case as well, and that the two pairs of $\pi/2$-fronts
$\chi=\pm\infty$ propagate in opposite directions.

\subsection{The effect of higher order terms}
\label{ho}
According to Eqn.~(\ref{chidot}) the asymptotic solutions just below
$\gamma_4=1/3$, the $\pi/2$-front pairs as $\vert\chi\vert\to\infty$,
are not smooth continuations of the stationary $\pi$-front at
$\gamma_4=1/3$ (the $\chi=0$ solution). This abrupt nature of the instability
is related to a degeneracy of solutions at $\gamma_4=1/3$. At this parameter 
value a whole family of solutions exists  describing
$\pi/2$-front pairs with distances $\vert
x_2-x_1\vert=2\vert\chi\vert$ ranging from zero to infinity. In the
nongradient case these pair solutions propagate at speeds given by
Eqn.~(\ref{zetadot_ng}).  The degeneracy of solutions is lifted by
higher order terms in Eqn.~(\ref{fcglB}).

\begin{figure}
\centering\includegraphics[width=3.0in]{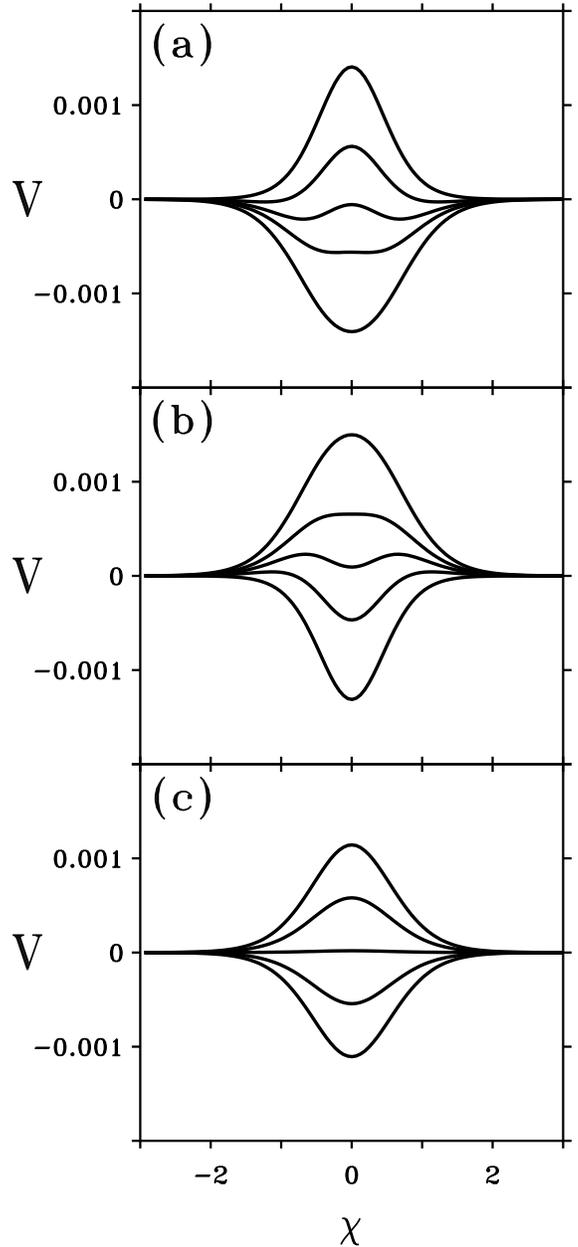}
\caption{The effects of the higher order term $\delta|B|^2
B_{xx}$ on the decomposition instability within the 4:1 resonance. 
The potential (\protect\ref{Vho}) deforms from a 
single well to a single barrier as $\gamma_4$ is decreased
past $\gamma_{4c}$.  In the intermediate range two scenarios 
are possible:
(a) For $\delta>0$, the $\chi=0$ solution loses stability in a pitchfork 
bifurcation at $\gamma_{4c}$ to a pair of solutions that move to $\pm\infty$.
Parameters: 
$\delta=1.0$, $\mu=0.01$, $\gamma_4=0.339,0.337,0.336,0.335,0.333$.
(b) For  $\delta<0$, the $\chi=0$ solution remains stable while the 
$\chi=\pm\infty$ solutions acquire stability and loses stability only below 
$\gamma_{4c}$.
Parameters: 
$\vert\delta\vert=1.0$, $\mu=0.01$, $\gamma_4=0.334,0.332,0.331,0.330,0.328$.
In both scenarios the deformations from a single well to a single barrier 
occur within a small range of $\protect\gamma_4$ of order $\mu\ll 1$. 
For comparison, an equivalent figure for the degenerate case ($\delta=0$) is 
shown in (c). The only intermediate form between a single well and a single 
barrier is a flat potential, $V=0$, ocurring at $\gamma=1/3$. 
}
\label{potential-higher}
\end{figure}

Consider the gradient version of the amplitude equation
\begin{equation}
B_t= B+\frac{1}{2}B_{xx}-|B|^2B +\gamma_4 {B^{*}}^3 +\mu
H(B,B^*;\partial_x) \,,
\label{hogl}
\end{equation}
where $ H(B,B^*;\partial_x)$ includes higher order terms like
$|B|^4B$, $|B|^2 B_{xx}$, etc..  The factor $\mu$ reflects the fact
that fifth order terms in the amplitude equation are smaller by factor
$\mu\ll 1$ than the lower order terms. The effect of these terms is
generally weak, but becomes important near $\gamma_4=1/3$. Consider,
for example, the effect of the term $\delta|B|^2
B_{xx}$. Eqs.~(\ref{UVeqn}) include now the contributions
\[
\frac{1}{2}\mu\delta (U^2+V^2)U_{xx}\quad {\rm and}\quad
\frac{1}{2}\mu\delta (U^2+V^2)V_{xx}\,,
\]
respectively. The corresponding contributions to Eqs.~(\ref{H1})
and~(\ref{H2}) are
\[ 
\frac{1}{2}\mu\delta \sigma_1\bigl[B_0^2(x-x_1)+B_0^2(x-x_2)\bigr]
B_0^{\prime\prime}(x-x_1)
\]
and
\[ 
\frac{1}{2}\mu\delta \sigma_2\bigl[B_0^2(x-x_1)+B_0^2(x-x_2)\bigr]
B_0^{\prime\prime}(x-x_2)\,.
\]
The equation for the order parameter will now read
\begin{equation}
\dot{\chi}=-\frac{27}{16}dJ(\chi )+\frac{9}{8}\delta\mu K(\chi)\,,
\label{hochidot}
\end{equation}
where
\begin{equation}
K(\chi )=\int_{-\infty }^{\infty }dz~\tanh {z}~{\rm sech}^{4}{z}~\tanh
^{2}(z+2\chi )\,.\label{K}
\end{equation}
The integral~(\ref{K}) is elementary but the expression is lengthy and
we do not display it here. The second term on the right hand side of
Eqn.~(\ref{hochidot}), whose origin is the fifth order term $\vert
B\vert^2B_{xx}$, cannot be neglected in a $\mu$-neighborhood of
$\gamma_4=1/3$. Depending on the sign of $\delta$ two scenarios are
possible as $\gamma_4$ is decreased. In both cases the $\chi=0$
($\pi$-front) solution is destabilized at
$\gamma_{4c}=1/3+8\mu\delta/21$. When $\delta>0$ the $\chi=0$ solution
is destabilized to a new pair of solutions, $\chi_\pm$, in a pitchfork
bifurcation. For $\vert\chi\vert\ll 1$ the new solutions assume the
approximate values, $\chi_\pm\approx
\pm\frac{\sqrt{21}}{4}\sqrt{1-d/d_c}$, where
$d_c=\gamma_{4c}-1/3$. When $\gamma_4$ is further decreased the two
stable solutions $\chi_\pm$ move to $\pm\infty$ on a $\gamma_4$ range
of order $\mu$. When $\delta<0$ bistability of the $\chi=0$ solution
and the $\chi=\pm\infty$ solutions first develop. As $\gamma_4$ is
further decreased the $\chi=0$ solution becomes metastable until it
loses completely its stability at $\gamma_{4c}$.
Fig.~\ref{potential-higher} shows the potential
\begin{equation}
V=\frac{9}{8}\int^\chi\Bigl[\frac{3}{2}dJ(z)-\delta\mu K(z)\Bigr]dz\,,
\label{Vho}
\end{equation}
associated with Eqn.~(\ref{hochidot}) for both scenarios.

The two scenarios are related by the symmetry $d\to -d$, $\delta\to
-\delta$, $t\to -t$ of Eqn.~(\ref{hochidot}).  The first scenario
($\delta>0$) amounts to a pitchfork bifurcation from a stable $\chi=0$
solution to a pair of stable $\chi_\pm$ solutions that move to
infinity as $\gamma_4$ is decreased. The second scenario ($\delta<0$)
amounts to a backward pitchfork bifurcation from an unstable $\chi=0$
solution to a pair of unstable $\chi_\pm$ solutions that move to
infinity as $\gamma_4$ is increased.

The higher order term $\vert B\vert^2B_{xx}$, and similarly other high
order terms, lift the degeneracy of the lower order system~(\ref{gl})
at $\gamma_4=1/3$. For $\delta>0$ and in a small $\gamma_4$ range of
order $\mu$ near 1/3 the instability becomes similar to the NIB
bifurcation in the 2:1 resonance. But apart from the behavior in this
small parameter range the overall behavior does not change: a
$\pi$-front decomposes into a pair of $\pi/2$-fronts as $\gamma_4$ is
decreased.


\section{$\Pi$-front instabilities in higher resonances} 
\begin{figure}
\centering\includegraphics[width=2.5in]{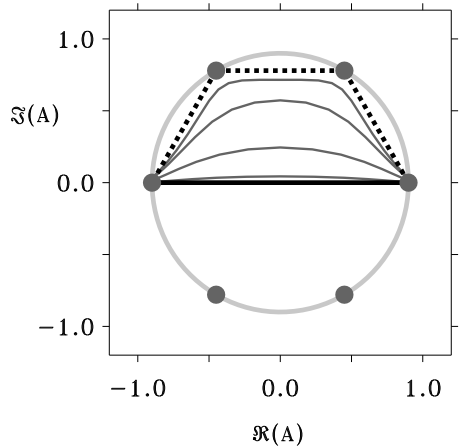}
\caption{Decomposition if a $\pi$ front into three $\pi/3$ fronts
in the 6:1 resonance band. Parameters in Eqn.~(\protect\ref{468eqn}):
 $\gamma_6=0.9$, $\mu_4=-1.0$, $\mu_6=-1.0$. 
All other parameters are zero. }
\label{6:1-phase}
\end{figure}

\begin{figure}
\centering\includegraphics[width=2.5in]{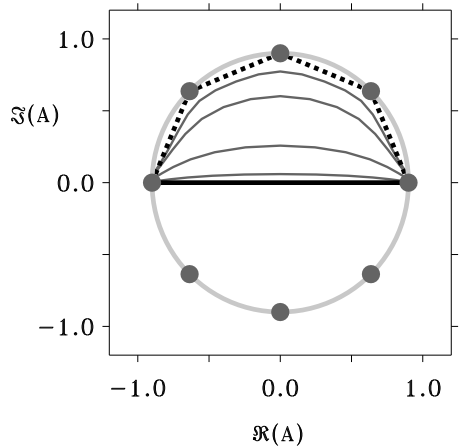}
\caption{
Decomposition if a $\pi$ front into four $\pi/4$ fronts in the 8:1
resonance band.  Parameters in Eqn.~(\protect\ref{468eqn}):
$\gamma_8=0.75$, $\mu_4=-0.5$, $\mu_6=-0.5$, $\mu_8=-1.0$.
All other parameters are zero.
}
\label{8:1-phase}
\end{figure}

\begin{figure}
\centering\includegraphics[width=3.0in]{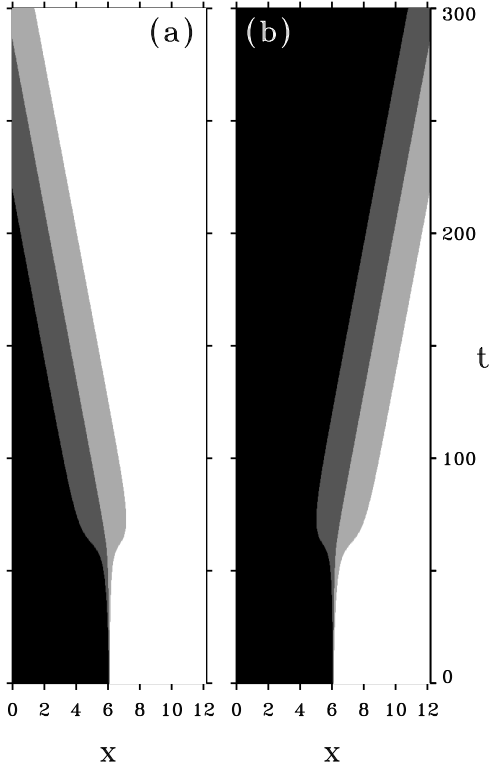}
\caption{
Decomposition if a $\pi$ front into three $\pi/3$ fronts in the 6:1
resonance band.  The figures show space-time plots of numerical
solutions of  Eqn.~(\protect\ref{468eqn}) with 
parameters $\gamma_6=0.9$, $\mu_4=-1.0$, $\mu_6=-1.0$, $\nu_0=0.1$.
All other parameters are zero.
}
\label{6:1}
\end{figure}

We have found
numerical evidence for the existence of similar $\pi$-front
instabilities within the 6:1 and 8:1 bands. These findings suggest the
following generalization: within the $2n$:1 band ($n>1$) a $\pi$-front
may lose stability by decomposing into $n$ $\pi/n$-fronts. Consider
the equation
\begin{eqnarray}
\label{468eqn} 
B_t&=& \frac{1}{2}B_{xx}+(1+i\nu_0)B \nonumber \\
&&\mbox{}+\mu_4|B|^2B+\mu_6|B|^4B +\mu_8|B|^6B\\
 &&\mbox{}+\gamma_4 {B^{*}}^3+\gamma_6 {B^{*}}^5+\gamma_8 {B^{*}}^7
   \nonumber \,.
\end{eqnarray}
The normal form equation up to seventh order contains many more terms.
Our purpose here, however, is just to demonstrate
the $\pi$-front instability for {\em some} parameter values pertaining
to the 6:1 and 8:1 bands. Fig.~\ref{6:1-phase} shows the decomposition
in the complex $B$ plane of a $\pi$-front within the 6:1 band into
three $\pi/3$-fronts.  Fig.~\ref{8:1-phase} shows the decomposition of
a $\pi$-front within the 8:1 band into four $\pi/4$-fronts.
Fig.~\ref{6:1} shows a space-time plot of the decomposition instability within the 
6:1 band. The initial unstable $\pi$-front decomposes into three  
$\pi/3$-fronts, traveling to the left or to the right depending on initial 
conditions. Along with this process two intermediate phase states appear 
between the original white and black phases. 

\section{Implications on pattern formation}
The $\pi$-front instability in the 4:1 band has a pronounced effect on
patterns.  Despite the coexistence of four uniform phase states and
the stability of $\pi/2$-fronts, asymptotic four-phase patterns appear
only below the $\pi$-front instability point $\gamma_4=1/3$. The
reason is the attractive interactions between $\pi/2$-fronts when 
$\gamma_4>1/3$ and the
collapse into $\pi$-fronts.  Thus, for $\gamma_4>1/3$ two-phase
patterns prevail. These patterns form standing waves since
$\pi$-fronts are stationary. For $\gamma_4<1/3$ the interaction
between $\pi/2$-fronts is repulsive and four-phase patterns
prevail. These patterns travel since $\pi/2$-fronts propagate.

\begin{figure}
\centering\includegraphics[width=2.5in]{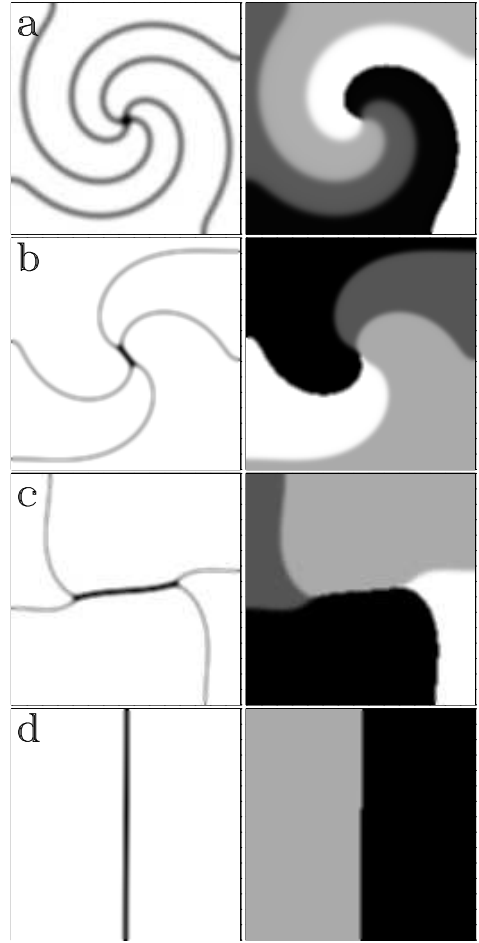}
\caption{Numerical solution of a two-dimensional version
of Eqn.~(\protect\ref{fcglB})
showing the collapse of a rotating four-phase spiral-wave into
a stationary two-phase pattern.
The left column is $|A|$ and the right column $\arg(A)$  in the $x-y$ plane.
(a) The initial four-phase spiral wave
(computed with $\gamma_4<1/3$).  
(b) The spiral core, a 4-point vertex, 
splits into two 3-point vertices connected by a $\pi$-front.
(c) A two-phase pattern develops as the 3-point vertices further separate.  
(d) The final stationary two-phase pattern.  
Parameters: $\gamma_4=0.6$, $\nu_0=0.1$, $\alpha=\beta=0$, 
$x=[0,64]$, $y=[0,64]$. 
}

\label{spiral}
\end{figure}

Fig.~\ref{spiral}$a$ shows a stably rotating four-phase spiral wave
for $\gamma_4<1/3$.
Figs.~\ref{spiral}$b,c,d$ show the collapse of this spiral wave into a
stationary two-phase pattern as $\gamma_4$ is increased past $1/3$.
The collapse begins at the spiral core where the $\pi/2$-front
interactions are the strongest. As pairs of $\pi/2$-fronts attract and
collapse into $\pi$-fronts, the core splits into two vertices that
propagate away from each other leaving behind a two-phase
pattern. 
%

\acknowledgments
We thank J. Guckenheimer and B. Krauskopf for helpful discussions.
This study was supported in part by grant  No 95-00112 from the US-Israel
Binational Science Foundation (BSF).

\bibliography{reaction}

\end{document}